\documentclass[12pt,letterpaper,oneside,onecolumn]{article}
\usepackage[T1]{fontenc}
\usepackage[dvips]{epsfig}

\makeatletter
\def\LyX{L\kern-.1667em\lower.25em\hbox{Y}\kern-.125emX\spacefactor1000}%
\newcommand{\lyxtitle}[1] {\thispagestyle{empty}
\global\@topnum\z@
\section*{\LARGE \centering \sffamily \bfseries \protect#1 }
}

\makeatother

\usepackage{epsfig}

\begin{document}

\hspace{11.cm} PM/98-09\\
\\

{\bfseries \LARGE \hfill{}SCALAR AND VECTOR MIXINGS\hfill{} \par}

{\bfseries \LARGE \hfill{}IN \( SU(2)\times U(1) \) MODELS\hfill{} \\
\\
\par}

{\bfseries \large \hfill{}M. Capdequi Peyranère\hfill{} \\
\par}

\hfill{}Laboratoire de Physique Mathématique et Théorique\hfill{}

\hfill{}CNRS- UMR 5825\hfill{} 

\hfill{}Place Eugène Bataillon\hfill{} 

\hfill{}34095 - Montpellier Cedex 05 (France)\hfill{} 

\vspace{2cm}

\em \hfill{}\bfseries Abstract\mdseries \hfill{} \em 

\vspace{0.1cm}

We study the possible mixings between gauge vector fields and scalar
fields through their self-energies, arising in models with two Higgs
doublets. We derive the relevant set of Schwinger-Dyson equations
and the Ward identities that compel the longitudinal parts of the
field propagators. Linear \( R_{\xi } \) gauge is used and the results are given
at all orders in perturbative theory, and some particular aspects
of the one loop case are stressed.

\vspace{4.5cm}

\centerline{To appear in  {\sf Int. J. Mod. Phys. A.}}

\newpage

\section{Introduction}

\renewcommand{\theequation}{\thesection.\arabic{equation}}

The Standard Model (SM) is nowadays tremendously confirmed by many
experimental results, even though there is not yet direct evidence
for scalar Higgs particle. Alternative models like the two Higgs doublet
model (THDM), especially the minimal supersymmetric standard model
(MSSM) are extensively studied, experimentally and theoretically as
well. It is now well settled that THDM suffer from large quantum effects,
and one loop corrections (and beyond) are really important to improve
the knowledge of such models (see for exemple\( ^{1} \) and references therein).
Particle mixings through quantum loops are common features of THDM
and SM and we will study some of them.

The mixing between the photon and the Z fields is well known and is
inherent to the Glashow -Salam - Weinberg (GSW) model. More precisely
, there is not only a mixing between the transverse parts of the propagators
of the photon and the Z but also a mixing between the longitudinal
parts of the photon and the Z propagators and the \( G^{0} \) neutral Goldstone
boson propagator. Moreover a mixing between the charged \( W^{\pm } \) fields and
the charged Goldstone \( G^{\pm } \) exists as well.

Actually these mixings come from the SU[2]\( * \)U[1] structure of the gauge
group and the Higgs mechanism, so in any model with this gauge group
and a symmetry breaking ``a la Higgs'' one gets such mixings. In the
GSW model, the mixings were studied thoroughly by Baulieu and Coquereaux\( ^{2} \)
. The aim of this paper is to extend their study to the THDM, where
the Higgs scalar sector is richer. Since we focus on the mixings coming
from the gauge group and the Higgs mechanism , our derivation is also
valid for the MSSM, without any change. Actually , we will examine
two mixings. The neutral set contains the photon , the Z , the Goldstone
\( G_{0} \) and the CP odd neutral Higgs scalar \( A_{0} \). In the charged set , there
are the gauge field \( W^{\pm } \), its associated Goldstone \( G^{\pm } \) and the Higgs field
\( H^{\pm } \).

In order to keep the results as general as possible , we will use
the BRS symmetry to find  the constraints on the two point Green's
functions. Indeed the BRS transformations of the fields are due to
the gauge structure and the gauge fixing term , which turn out to
be the same in all the models we consider. For the same reason, we
carry out a derivation valid at all orders in perturbation theory,
that gives as a by-product the one loop results and shows the physical
differences between first order and all orders. 

For the sake of generality, because there is a large variety of possible
scenarii, we don't at all study the renormalization of the two point
functions, mixed or not mixed. Indeed, one can choose different renormalization
conditions, one can renormalize before or after breaking of the SU[2]\( * \)U[1]
gauge symmetry, one can choose a renormalization scheme where the
gauge vector mixing ( photon-Z ) is present or absent, and , in the
case of the MSSM, the dimensional reduction could be more adapted
than the dimensional regularization. Moreover, if one wishes that
the constraints we derive using the BRS symmetry be respected before
and after the renormalization procedure, it will be necessary to choose
it in such a way the gauge fixing lagrangian be renormalization invariant,
that it is always possible\( ^{3} \) . Finally, we will only assume that the
Green functions we use are regularized with a gauge invariant method.

This paper contains three sections. The first one is mainly devoted
to display the tools we will use in the other sections. We begin by
our notations that we wish natural and compact in such a way that
the Schwinger -Dyson equations are comprehensive and clear. We also
recall some details on the BRS transformations, useful to derive the
constraints induced by the BRS symmetry on the various two point Green
functions. 

In the second section we study the charged case, where we have to
take into account three fields, six two point functions and one constraint.
It is in some sense a short introduction to the neutral case worked
out in the third section, a much more complex case since we have to
handle four fields , ten two point functions and three constraints.

\section{Preliminaries}

\setcounter{equation}{0}

\subsection{Quantum numbers and mixings.}

In the THDM (or MSSM), the C and P quantum numbers of the Higgs or
Goldstone scalar fields and the vector fields are well known\( ^{4} \). They
are summarized in the following table:

\vspace{0.30cm}
{\centering \begin{tabular}{|c|c|c|}
\hline 
field&C \& P conserved&CP conserved\\
\hline 
\hline 
\( \gamma  \)&\( -- \)&\( -- \)\\
\hline 
\( Z \)&\( -- \)&\( --;++ \)\\
\hline 
\( G_{0} \)&\( +- \)&\( +-;-+ \)\\
\hline 
\( A_{0} \)&\( +- \)&\( +-;-+ \)\\
\hline 
\( H_{0} \)&\( ++ \)&\( ++;-- \)\\
\hline 
\( h_{0} \)&\( ++ \)&\( ++;-- \)\\
\hline 
\( W^{\pm } \)&\( - \)&\( -;+ \)\\
\hline 
\( G^{\pm } \)&\( + \)&\( +;- \)\\
\hline 
\( H^{\pm } \)&\( + \)&\( +;- \)\\
\hline 
\end{tabular}\par}
\vspace{0.30cm}

The second column concerns the bosonic interactions when charge conjugation
and spacial parity are separately conserved, while the third column
includes the fermionic sector which is CP invariant. In the case of
charged fields, C is not a good quantum number, so it has been discarded.
If a neutral vector field \( V^{\mu } \) has the \( 1^{PC} \) quantum number assignments ,
then its derivative \( \partial _{\mu }V^{\mu } \) is a neutral scalar with the \( 0^{-PC} \) assignments.
That allows some mixings between scalars and vectors via their longitudinal
parts but also it forbids the mixing between the CP even Higgs scalar
fields and the photon or the Z. This is the true reason why the Standard
Model Higgs scalar field \( H \) ( which has the same quantum numbers as
the THDM \( h_{0} \) and \( H_{0} \) ) does not couple to the neutral vector gauge fields.
In any case, one finds that the neutral mixing set contains the four
first particles of this table and the charged set involves all the
charged particles.

\subsection{Notations}

We now define our notations for the two point functions. They are
slightly different from the notations of reference\( ^{2} \).

First we denote by \( L \) and \( T \) the longitudinal and transverse projectors
:

\[
L^{\mu \nu }=k^{\mu }k^{\nu }/k^{2}\, \mbox {and}\, T^{\mu \nu }=g^{\mu \nu }-L^{\mu \nu }\, ,\]

\noindent in such a way that the free propagator of a gauge vector
field reads

\[
D^{\mu \nu }=-i(T^{\mu \nu }/(k^{2}-m^{2})+\xi \, L^{\mu \nu }/(k^{2}-\xi m^{2}))\]

\noindent where \( \xi  \) is the gauge fixing parameter. To be complete, the
free propagator of a massive scalar field is \( D=i/(k^{2}-m^{2}) \) while the propagator
of a Goldstone field is \( D=i/(k^{2}-\xi m^{2}) \).

We will note \( G_{ij} \) the complete two point Green function and \( P_{ij} \) the one
particule ireducible (1PI) two point Green function, where i and j
are field indices:

\[
G_{ij}=<0|i(x)j(y)|0>.\]

\noindent In our definition of the 1PI two point Green functions,
the tadpoles are allowed since we cope with expressions before renormalization.
We can now introduce the transverse and longitudinal parts of these
Green functions ( in momentum space ), where M is a totally arbitrary
mass dimensioned parameter:

\[
G^{\mu \nu }_{ij}=-i(G_{ij}^{T}T^{\mu \nu }+G^{L}_{ij}L^{\mu \nu })\, \, \, ;\, \, \, P^{\mu \nu }_{ij}=+i(P^{T}_{ij}T^{\mu \nu }+P^{L}_{ij}L^{\mu \nu })\]

\[
G^{\mu }_{ij=}-ik^{\mu }G^{L}_{ij}/M\, \, \, ;\, \, \, P^{\mu }_{ij}=+ik^{\mu }P^{L}_{ij}/M\]

\[
G_{ij}=-iG^{L}_{ij}\, \, \, ;\, \, \, P_{ij}=+iP^{L}_{ij}\]

In the previous formulae , two Lorentz indices mean a vector-vector
mixing, one Lorentz index means a vector-scalar mixing and no Lorentz
index means a scalar-scalar mixing. With these formulae, the \( G \) functions
( like the \( D \) functions ) and \( P \) functions have respectively the canonical
dimension -2 and + 2.

\subsection{The Schwinger - Dyson equations}

In pure QED the Schwinger -Dyson equation can be proved analytically,
or diagramatically by ``sticking'' the product \( PD \) to the geometrical
expansion of \( G \) in terms of \( D \) and \( P \). In both cases one gets the well
known result: \( G=D+DPG \) ,whose diagrammatic form is:

$$\epsfig{file=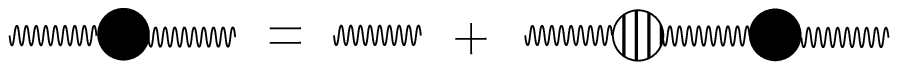} $$

\noindent This pure QED equation can be easily extended when mixings
arise. Using a matrix formalism, one readily finds the general equation
\( G=D+DPG \), where \( G \) , \( D \) and \( P \) are now matrices. If a free field propagator
is generically painted as a solid line, and a complete ( 1PI ) Green
function is painted as a black ( grid ) disk between two solide lines
, then the pictorial form of this matrix equation is as follows: 

$$\epsfig{file=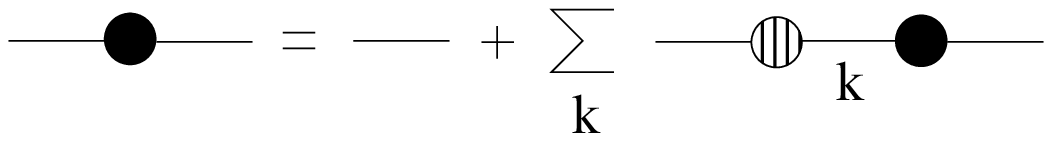}$$

\noindent As it is possible to get vector-vector , vector-scalar and
scalar-scalar mixings , we can display three equations , where \( v \) (\( s \))
is the set of all the possible vector (scalar) fields:

\[
G^{\mu \nu }_{V_{i}V_{j}}=D^{\mu \nu }_{V_{i}}\delta _{ij}+\sum _{m\in v}D^{\mu \rho }_{V_{i}}P_{V_{i}V_{m}}^{\rho \sigma }G_{V_{m}V_{j}}^{\sigma \nu }+\sum _{n\in s}D^{\mu \rho }_{V_{i}}P^{\rho }_{V_{i}S_{n}}G^{\nu }_{S_{n}V_{j}}\, ;\]

\[
G^{\mu }_{V_{i}S_{l}}=\sum _{m\in v}D^{\mu \rho }_{V_{i}}P^{\rho \sigma }_{V_{i}V_{m}}G^{\sigma }_{V_{m}S_{l}}+\sum _{n\in s}D^{\mu \rho }_{V_{i}}P^{\rho }_{V_{i}S_{n}}G_{S_{n}S_{l}}\, ;\]

\[
G_{S_{k}S_{l}}=D_{S_{k}}\delta _{kl}+\sum _{m\in v}D_{S_{k}}P^{\rho }_{S_{k}V_{m}}G^{\rho }_{V_{m}S_{l}}+\sum _{n\in s}D_{S_{k}}P_{S_{k}S_{n}}G_{S_{n}S_{l}}\, .\]

Introducing the transverse and longitudinal Green functions, one obtains
four coupled equations, one tranverse and three longitudinal: 
\begin{equation}
\label{2.1}
G^{T}_{V_{i}V_{j}}=(\delta _{ij}+\sum _{m\in v}P^{T}_{V_{i}V_{m}}G^{T}_{V_{m}V_{j}})/(k^{2}-m^{2}_{i})
\end{equation}

\begin{equation}
\label{2.2}
G^{L}_{V_{i}V_{j}}=(\delta _{ij}+\sum _{m\in v}P^{L}_{V_{i}V_{m}}G^{L}_{V_{m}V_{j}}+k^{2}\sum _{n\in s}P^{L}_{V_{i}S_{n}}G^{L}_{S_{n}V_{j}}/M^{2})/(k^{2}-\xi m_{i}^{2})
\end{equation}

\begin{equation}
\label{2.3}
G^{L}_{V_{i}S_{l}}=\xi (\sum _{m\in v}P^{L}_{V_{i}V_{m}}G^{L}_{V_{m}S_{l}}+\sum _{n\in s}P^{L}_{V_{i}S_{n}}G^{L}_{S_{n}S_{l}})/(k^{2}-\xi m^{2})
\end{equation}

\begin{equation}
\label{2.4}
G^{L}_{S_{k}S_{l}}=-(\delta _{kl}+k^{2}\sum _{m\in v}P^{L}_{S_{k}V_{m}}G^{L}_{V_{m}S_{l}}/M^{2}+\sum _{n\in s}P^{L}_{S_{k}S_{n}}G^{L}_{S_{n}S_{l}})/(k^{2}-m^{2}_{k})
\end{equation}

\noindent In the last equation, if \( S_{k} \) is a Goldstone boson, \( k^{2}-m^{2}_{k} \) must
be changed in \( k^{2}-\xi m^{2}_{k} \). Of course, \( G \) and \( P \) are symmetric functions in their
arguments. These four equations are the most general Schwinger -Dyson
equations for mixings between spin zero and spin one particles.  Their
solutions give the complete two point Green functions \( G \) in terms of
the 1PI Green functions \( P \).

\subsection{Ward identities.}

Since the THDM is a gauge model, Green functions are not independent
and they are constrained by the BRS symmetry , the quantum level symmetry
that encompasses the gauge symmetry valid at the classical level.
Using the BRS transformations, one gets identities between the complete
Green functions \( G \), which in turn give identities between the 1PI Green
functions \( P \), when solved the above Schwinger -Dyson equations.

First, one considers the following two point Green function :

\[
<0|\eta ^{*}_{i}(x)F_{j}(y)|0>,\]

\noindent where \( F_{j} \) is the gauge fixing term associated to the gauge
vector field \( V_{j} \), and \( \eta _{i} \) is the Faddev-Popov ghost associated to the
gauge vector field \( V_{i} \). By fermionic antisymmetry , this Green function
is null, and so its BRS transform. Since the BRS transform of an anti-ghost
\( \eta _{i}^{*} \) is the gauge fixing term \( F_{i} \) and the BRS transform of a gauge fixing
term is proportional to the equation of motion of the corresponding
anti-ghost , one gets\( ^{2,3,5} \)

\begin{equation}
\label{2.5}
<0|F_{i}(x)F_{j}(y)|0>=\delta _{ij}\delta (x-y)
\end{equation}

This equation gives us the first set of Ward identities between the
two point Green functions \( G \) . Since they only mix gauge vector fields
and Goldstone fields, these identities are valid in the GSW model
and in any model with a larger Higgs sector as well. To be precise
we will use in this paper the definitions of Bohm et al.\( ^{5} \) for the
gauge vector fields and the gauge fixing terms, i.e.:

\[
A_{\mu }=-sin\theta _{w}W^{3}_{\mu }+cos\vartheta _{w}B_{\mu };Z_{\mu }=cos\theta _{w}W^{3}_{\mu }+sin\theta _{w}B_{\mu }\, \, \, ;\]

\[
W^{\pm }_{\mu }=(W^{1}_{\mu }\mp iW^{2}_{\mu })/\sqrt{2}\, \, \, ;\]

\[
F_{p}=\partial _{\mu }A^{\mu }/\sqrt{\xi };F_{z}=(\partial _{\mu }Z^{\mu }-\xi m_{z}G_{0})/\sqrt{\xi }\, \, \, ;\]

\[
F^{\pm }=(\partial ^{\mu }W^{\pm }_{\mu }\mp i\xi m_{w}G^{\pm })/\sqrt{\xi }.\]

A second set of Ward identities can be obtained with the BRS variation
of the following Green function: \( <0|\eta ^{*}_{i}(x)S_{j}(y)|0> \) when \( S_{j} \) is a Higgs scalar field.
It is also a null function and its BRS variation reads:

\begin{equation}
\label{2.6}
<0|F_{i}(x)S_{j}(y)|0>-<0|\eta ^{*}_{i}(x)\delta _{BRS}S_{j}(y)|0>=0
\end{equation}

\noindent Such a formula cannot give simple relations between the
\( G \) functions studied here, since the second term on the left side is
a composite operator with an external ghost field. Because the fermionic
ghost number is conserved, only this \( \eta ^{*}_{i} \) can give a pole in the second
term. Hence, multiplying by the inverse propagator of the scalar field
\( S_{j} \) and putting it on its mass shell ( both operations noticed ``S.o.s''
), the composite operator vanishes, and we just get identities between
the \( G \) Green functions we consider here: 
\begin{equation}
\label{2.7}
<0|F_{i}(x)S_{j}(y)|0>_{S.o.s}=0
\end{equation}
 Moreover, when \( F_{i} \) is the
photon or the \( Z \) gauge fixing term and \( S_{j} \) is the CP-odd neutral Higgs
\( A_{0} \), equation (1.4) gives us an useful result at one loop order\( ^{6} \). Indeed
the BRS transform of the \( A_{0} \) contains fields ( scalars and F.P ghosts
) in such a way that no tree level coupling exists with the \( \eta _{\gamma } \) and
\( \eta _{z} \) ghosts . Hence at one loop order ( ``o.l.o'' index ) we get the following
equations : 
\begin{equation}
\label{2.8}
[k^{2}G^{L}_{ph}]_{o.l.o}=<0|\partial _{\mu }A^{\mu }(x)A_{0}(y)|0>_{o.l.o}=0
\end{equation}
and 
\begin{equation}
\label{2.9}
[k^{2}G^{L}_{zh}+i\xi Mm_{z}G^{L}_{gh}]_{o.l.o}=<0|(\partial _{\mu }Z^{\mu }(x)-\xi m_{z}G_{0}(x))A_{0}(y)|0>_{o.l.o}=0
\end{equation}

\noindent which are true independently of any mass or energy prescription.

It is to be noted that this second set of Ward identities (2.6), (2.7),
(2.8) and (2.9) is of interest only for the extensions of the GSW
model, as the THDM or MSSM, where the Higgs sector is rich enough
to contains a neutral CP odd \( A_{0} \) and a charged scalar \( H^{\pm } \), together with
correct quantum numbers. It is also to be stressed that the identities
due to the BRS symmetry are not in general renormalization invariant;
but the way we derive the identities (2.5), (2.7), (2.8) and (2.9)
assures they can be preserved by renormalization, as we said in the
introduction.

\section{Charged fields}

\setcounter{equation}{0}

The charged gauge boson \( W^{\pm } \) can be mixed with the Goldstone \( G^{\pm } \) and the
Higgs \( H^{\pm } \), but only by its longitudinal part . For the transverse part
, there is no mixing at all and the equation (2.1), which is very
similar to the photon one in pure QED , reads , omitting the \( \pm  \) index
:

\[
G^{T}_{ww}=(1+P^{T}_{ww}G^{T}_{ww})/(k^{2}-m^{2}_{w})\, ;\]

\noindent the solution is obvious:

\[
G^{T}_{ww}=1/(k^{2}-m^{2}_{w}-P^{T}_{ww}).\]

Actually the interest is in the six coupled longitudinal equations
coming from (2.2), (2.3) and (2.4) where we also have discarded the
L index :

\[
G_{ww}=\xi [1+P_{ww}G_{wg}+k^{2}(P_{wg}G_{gw}+P_{wh}G_{hw})/M^{2}]/(k^{2}-\xi m_{w}^{2})\, \, \, ;\]

\[
G_{wg}=\xi [P_{ww}G_{wg}+P_{wg}G_{gg}+P_{wh}G_{hg}]/(k^{2}-\xi m^{2}_{w})\, \, \, ;\]

\[
G_{wh}=\xi [P_{ww}G_{wh}+P_{wg}G_{gh}+P_{wh}G_{hh}]/(k^{2}-\xi m^{2}_{w})\, \, \, ;\]

\[
G_{gg}=-[1+k^{2}P_{gw}G_{wg}/M^{2}+P_{gh}G_{hg}+P_{gg}G_{gg}]/(k^{2}-\xi m^{2}_{w})\, \, \, ;\]

\[
G_{gh}=-[k^{2}P_{gw}G_{wh}/M^{2}+P_{gg}G_{gh}+P_{gh}G_{hh}]/(k^{2}-\xi m^{2}_{w})\, \, \, ;\]

\[
G_{hh}=-[1+k^{2}P_{hw}G_{wh}/M^{2}+P_{hg}G_{gh}+P_{hh}G_{hh}]/(k^{2}-m^{2}_{h})\, .\]

\noindent A solution of this linear system is rather easy to find
with an analytic software but the solution suffers potentially large
simplifications and it is highly preferable to ``prepare'' the system
as follows :

\[
\begin{array}{lllll}
\left( \begin{array}{cccccc}
a_{11} & a_{12} & 0 & a_{14} & 0 & 0\\
0 & a_{11} & a_{23} & 0 & a_{25} & 0\\
0 & a_{12} & a_{33} & 0 & a_{35} & 0\\
0 & 0 & 0 & a_{11} & a_{23} & a_{25}\\
0 & 0 & 0 & a_{12} & a_{33} & a_{35}\\
0 & 0 & 0 & a_{14} & a_{35} & a_{66}
\end{array}
\right)  & . & \left( \begin{array}{c}
G_{ww}\\
G_{wg}\\
G_{gg}\\
G_{wh}\\
G_{gh}\\
G_{hh}
\end{array}
\right)  & =\xi  & \left( \begin{array}{c}
1\\
0\\
1\\
0\\
0\\
1
\end{array}
\right) 
\end{array}
\]

\noindent with the notations :

\[
a_{11}=k^{2}-\xi m_{w}^{2}-\xi P_{ww}\, \, \, ;\, \, \, a_{12}=k^{2}a_{23}/M^{2};a_{14}=k^{2}a_{25}/M^{2};\]

\[
a_{23}=-\xi P_{wg}\, ;\, a_{25}=-\xi P_{wh}\, \, ;\]

\[
a_{33}=-(k^{2}-\xi m^{2}_{w}+P_{gg})\, ;\, a_{35}=-\xi P_{gh}\, \, ;\]

\[
a_{66}=-\xi (k^{2}-m^{2}_{h}+P_{hh})\, .\]

\noindent Now it is easy to obtain the most compact solution of the
longitudinal charged mixing:

\[
G_{ww}=\xi M^{2}[-P_{gh}^{2}+(k^{2}-\xi m_{w}^{2}+P_{gg})(k^{2}-m^{2}_{h}+P_{hh})]/D,\]

\[
G_{wg}=\xi M^{2}[P_{gh}P_{wh}-P_{wg}(k^{2}-m_{h}^{2}+P_{hh})]/D,\]

\[
G_{gg}=-[\xi k^{2}P_{wh}^{2}+M^{2}(k^{2}-m_{h}^{2}+P_{hh})(k^{2}-\xi m^{2}_{w}-\xi P_{ww})]/D,\]

\[
G_{wh}=\xi M^{2}[P_{gh}P_{wg}-P_{wh}(k^{2}-\xi m^{2}_{w}-\xi P_{ww})]/D,\]

\[
G_{gh}=[\xi k^{2}P_{wg}P_{wh}+M^{2}P_{gh}(k^{2}-\xi m_{w}^{2}-\xi P_{ww})]/D,\]

\begin{equation}
\label{3.1}
G_{hh}=-[\xi k^{2}P_{wg}^{2}+M^{2}(k^{2}-\xi m_{w}^{2}+P_{gg})(k^{2}-\xi m_{w}^{2}-\xi P_{ww})]/D,
\end{equation}

\noindent where the denominator \( D \) is :

\[
D=M^{2}P_{gh}^{2}(k^{2}-\xi m^{2}_{w}-\xi P_{ww})\]

\[
+\xi k^{2}[P_{wg}^{2}(k^{2}-m_{h}^{2}+P_{hh})+P_{wh}^{2}(k^{2}-\xi m^{2}_{w}+P_{gg})-2P_{gh}P_{wg}P_{wh}]\]

\begin{equation}
\label{3.2}
+M^{2}(k^{2}-\xi m^{2}_{w}+P_{gg})(k^{2}-m^{2}_{h}+P_{hh})(k^{2}-\xi m^{2}_{w}-\xi P_{ww}).
\end{equation}

At one loop order, it remains :

\[
G_{ww}=\xi (k^{2}-\xi m^{2}_{w}+\xi P_{ww})/(k^{2}-\xi m^{2}_{w})^{2},\]

\[
G_{wg}=-\xi P_{wg}/(k^{2}-\xi m^{2}_{w})^{2},\]

\[
G_{gg}=-(k^{2}-\xi m^{2}_{w}-P_{gg})/(k^{2}-\xi m^{2}_{w})^{2},\]

\[
G_{wh}=-\xi P_{wh}/[(k^{2}-m^{2}_{h})(k^{2}-\xi m^{2}_{w})],\]

\[
G_{gh}=P_{gh}/[(k^{2}-m^{2}_{h})(k^{2}-\xi m^{2}_{w})],\]

\begin{equation}
\label{3.3}
G_{hh}=-(k^{2}-m^{2}_{h}-P_{hh})/(k^{2}-m^{2}_{h})^{2}.
\end{equation}

\noindent As a consequence of this one loop result, we see that the
first three Green functions \( G_{ww} \) , \( G_{wg} \) and \( G_{gg} \) become independent of \( P_{gh} \), \( P_{wh} \)
and \( P_{hh} \) . Therefore, within the GSW model, these three longitudinal
Green functions (at one loop order) have exactly the same form as
we have got.

The relevant Ward identity (2.5) in the charged sector is:

\[
<0|F^{\pm }(x)F^{\pm }(y)|0>=\delta _{\pm \pm }\delta (x-y)\]

\noindent and reads in the momentum space :

\begin{equation}
\label{3.4}
k^{2}G_{ww}\pm 2\xi m_{w}k^{2}G_{wg}/M+\xi ^{2}m^{2}_{w}G_{gg}=\xi 
\end{equation}

\noindent the \( \pm  \) sign depending on the field \( W^{\pm } \). When we replace in
this equation the above solutions, we get the following constraint
between the various \( P \) Green functions:

\[
2k^{2}P_{gh}P_{wg}P_{wh}-k^{2}P_{gg}P_{wh}^{2}-(Mm_{w}P_{gh}\mp k^{2}P_{wh})^{2}-M^{2}P_{ww}P_{gh}^{2}\]

\begin{equation}
\label{3.5}
+(k^{2}-m^{2}_{h}+P_{hh})[M^{2}(k^{2}+P_{gg})(m^{2}_{w}+P_{ww})-k^{2}(P_{wg}\pm Mm_{w})]=0.
\end{equation}

It is possible to simplify the denominator in the solutions of the
previous system thanks to the Ward identity; then we get the following
formula :

\[
D=-k^{2}(MP_{gh}\mp \xi m_{w}P_{wh})^{2}\]

\[
+M(k^{2}-m^{2}_{h}+P_{hh})(M(k^{2}-\xi m^{2}_{w})^{2}+Mk^{2}P_{gg}\mp 2\xi k^{2}m_{w}P_{wg}+\xi ^{2}Mm^{2}_{w}P_{ww}).\]

\noindent Of course the Ward identity (3.4) will be automatically
fulfiled if one uses such a denominator.

Again , giving up the scalar Higgs \( H^{\pm } \) as in the GSW model , the Ward
identity (3.5) is simpler and reads:

\begin{equation}
\label{3.6}
k^{2}(Mm_{w}\pm P_{wg})^{2}=M^{2}(k^{2}+P_{gg})(m^{2}_{w}+P_{ww})
\end{equation}

\noindent This equation is quite similar to the equation 3.12 in \( ^{2} \).
It is worth noting this equation is almost what we get from the full
equation at one loop order, since we just have to discard \( P^{2} \) terms
to end with:

\begin{equation}
\label{3.7}
Mm^{2}_{w}P_{gg}\mp 2k^{2}m_{w}P_{wg}+k^{2}MP_{gg}=0
\end{equation}

\noindent So this constraint on the self-energies is at one loop order
the same in both models, GSW or THDM. In other words , at one loop
order, the charged Higgs \( H^{\pm } \) decouples in this constraint, which could
be expected from the one loop solution of the system (formulae (3.3)).

The second constraint we have in the charged sector follows directly
from the equation (2.6) and we get the relation between longitudinal
self -energies, with an on shell Higgs field:
\[
M(m^{2}_{h}-\xi m^{2}_{w})(m^{2}_{h}P_{wh}\pm Mm_{w}P_{gh})+\]

\[
MP_{gh}(\xi Mm_{w}P_{ww}\mp m^{2}_{h}P_{wg})+m^{2}_{h}P_{wh}(\xi m_{w}P_{wg}\mp MP_{gg})=0.\]
At one loop order the
previous equations obviously reads\( ^{7} \) :

\[
m^{2}_{h}P_{wh}\pm Mm_{w}P_{gh}=0\]

\section{Neutral fields}

\setcounter{equation}{0}

Likewise the third section , we begin by the transverse neutral case
where the mixing between the photon and the neutral boson Z appears;
we have three coupled equations ( 2.1 ), omitting the T label:

\[
G_{pp}=(1+P_{pp}G_{pp}+P_{pz}G_{pz})/k^{2};\]

\[
G_{pz}=(P_{pp}G_{pz}+P_{pz}G_{zz})/k^{2};\]

\[
G_{zz}=(1+P_{pz}G_{pz}+P_{zz}G_{zz})/(k^{2}-m^{2}_{z}).\]

The solutions are ;

\[
G_{pp}=(k^{2}-m^{2}_{z}-P_{zz})/[(k^{2}-P_{pp})(k^{2}-m_{z}^{2}-P_{zz})-P^{2}_{pz}];\]

\[
G_{pz}=P_{pz}/[(k^{2}-P_{pp})(k^{2}-m^{2}_{z}-P_{zz})-P^{2}_{pz}];\]

\[
G_{zz}=(k^{2}-P_{pp})/[(k^{2}-P_{pp})(k^{2}-m^{2}_{z}-P_{zz})-P^{2}_{pz}].\]

The really involved part is the longitudinal case where we have to
deal with ten coupled equations ; indeed we have to take into account
the photon , the neutral gauge vector boson Z, the Goldstone scalar
\( G_{0} \) and the CP odd scalar Higgs \( A_{0} \) , so we get 10 (= 4+3+2+1) possible
\( G \) and \( P \) Green functions. To be short, we will not fully write these
equations like in the third section , since they are easy to derive
from the general Schwinger - Dyson relations ( 2.2 ). 

These complete longitudinal Green functions (omitting now the L label)
verify the three following Ward identities derived from equation (2.5)
:

\begin{equation}
\label{4.1}
k^{2}G_{pp}=\xi 
\end{equation}

\begin{equation}
\label{4.2}
MG_{pz}=i\xi m_{z}G_{pg}
\end{equation}

\begin{equation}
\label{4.3}
k^{2}MG_{zz}-2i\xi m_{z}k^{2}G_{zg}+\xi ^{2}m^{2}_{z}MG_{gg}=\xi M
\end{equation}

\noindent and we will solve our problem using these constraints. If
one try to solve this 10x10 system by brute force one gets a huge
result. Once again, we prefer to solve it in the following compact
form :

\[
\begin{array}{lllll}
\left( \begin{array}{cccccccccc}
a_{11} & a_{12} & 0 & a_{14} & 0 & 0 & a_{17} & 0 & 0 & 0\\
0 & a_{11} & a_{12} & 0 & a_{14} & 0 & 0 & a_{17} & 0 & 0\\
0 & a_{12} & a_{33} & 0 & a_{35} & 0 & 0 & a_{38} & 0 & 0\\
0 & 0 & 0 & a_{11} & a_{12} & a_{46} & 0 & 0 & a_{49} & 0\\
0 & 0 & 0 & a_{12} & a_{33} & a_{56} & 0 & 0 & a_{59} & 0\\
0 & 0 & 0 & a_{14} & a_{35} & a_{66} & 0 & 0 & a_{69} & 0\\
0 & 0 & 0 & 0 & 0 & 0 & a_{11} & a_{12} & a_{46} & a_{49}\\
0 & 0 & 0 & 0 & 0 & 0 & a_{12} & a_{33} & a_{56} & a_{59}\\
0 & 0 & 0 & 0 & 0 & 0 & a_{14} & a_{35} & a_{66} & a_{69}\\
0 & 0 & 0 & 0 & 0 & 0 & a_{17} & a_{38} & a_{69} & a_{00}
\end{array}
\right)  & . & \left( \begin{array}{c}
G_{pp}\\
G_{pz}\\
G_{zz}\\
G_{pg}\\
G_{zg}\\
G_{gg}\\
G_{ph}\\
G_{zh}\\
G_{gh}\\
G_{hh}
\end{array}
\right)  & =\xi  & \left( \begin{array}{c}
1\\
0\\
1\\
0\\
0\\
1\\
0\\
0\\
0\\
1
\end{array}
\right) 
\end{array}
\]

where the coefficients \( a_{ij} \) read as follows:\\

\( a_{14}=k^{2}a_{46}/M^{2};a_{17}=k^{2}a_{49}/M^{2};a_{35}=k^{2}a_{56}/M^{2};a_{38}=k^{2}a_{59}/M^{2}; \)

\vspace{0.3cm}

\( a_{11}=k^{2}-\xi P_{pp};a_{12}=-\xi P_{pz};a_{33}=k^{2}-\xi m^{2}_{z}-\xi P_{zz}; \)

\vspace{0.3cm}

\( a_{46}=-\xi P_{pg};a_{49}=-\xi P_{Ph};a_{56}=-\xi P_{zg;};a_{59}=-\xi P_{zh}; \)

\vspace{0.3cm}

\( a_{66}=-\xi (k^{2}-\xi m^{2}_{z}+P_{gg});a_{69}=-\xi P_{gh};a_{00}=-\xi (k^{2}-m^{2}_{h}+P_{hh}). \)\\

Since we have to deal with a 10x10 matrix, we naively expect that
the determinant is a polynomial of degree 10 in \( a_{ij} \), but if we smartly
perform the derivation , we get solutions as:
\[
G_{ij}=N_{ij}/\delta ,\]
 \noindent where \( N_{ij} \) are
polynomials of degree 3 while \( \delta  \) is a polynomial of degree 4, which
is much better . Now using the first constraint , i.e. the photon
Ward identity ( 4.1 ) , \( N_{pp}k^{2}=\xi \delta  \), we can replace \( \delta  \) by a polynomial of degree
3. Then we obtain solutions such that \( G_{pp} \) = \( \xi /k^{2} \) and the other functions
\( G_{ij} \) = \( n_{ij} \)/ \( k^{2}d \), where now \( d \) and \( n_{ij} \) are both polynomials of third degree .
Of course we still have to rewrite everything in terms of the \( P_{ij} \) functions
to end up with the final result:
\[
G_{pz}=-\xi ^{2}[M^{2}P_{pz}(k^{2}-\xi m^{2}_{z}+P_{gg})(k^{2}-m^{2}_{h}+P_{hh})-k^{2}P_{pg}P_{zg}(k^{2}-m^{2}_{h}+P_{hh})+\]

\[
k^{2}P_{gh}(P_{ph}P_{zg}+P_{pg}P_{zh})-k^{2}P_{ph}P_{zh}(k^{2}-\xi m^{2}_{z}+P_{gg})-M^{2}P^{2}_{gh}P_{pz}]/k^{2}d.\]

\[
G_{zz}=-\xi [\xi k^{2}P_{pg}^{2}(k^{2}-m^{2}_{h}+P_{hh})-\xi k^{2}(2P_{gh}P_{pg}P_{ph}-P_{ph}^{2}(k^{2}-\xi m_{z}^{2}+P_{gg}))-\]

\[
M^{2}(k^{2}-\xi P_{pp})(P^{2}_{gh}-(k^{2}-m^{2}_{h}+P_{hh})(k^{2}-\xi m^{2}_{z}+P_{gg}))]/k^{2}d.\]

\[
G_{pg}=\xi [\xi M^{2}P_{pz}(P_{zg}(k^{2}-m^{2}_{h}+P_{hh})-P_{gh}P_{zh})-\xi k^{2}(P_{ph}P_{zg}P_{zh}-P_{pg}P_{zh}^{2})-\]

\[
M^{2}(k^{2}-\xi m_{z}^{2}-\xi P_{zz})(P_{gh}P_{ph}-P_{pg}(k^{2}-m^{2}_{h}+P_{hh}))]/k^{2}d.\]

\[
G_{zg}=-\xi [\xi M^{2}P_{pz}(P_{gh}P_{ph}-(k^{2}-m^{2}_{h}+P_{hh})P_{pg}-\xi k^{2}P_{ph}(P_{ph}P_{zg}-P_{pg}P_{zh})+\]

\[
M^{2}(k^{2}-\xi P_{pp})(P_{gh}P_{zh}-P_{zg}(k^{2}-m^{2}_{h}+P_{hh}))]/k^{2}d.\]

\[
G_{gg}=[\xi ^{2}P_{pz}(2k^{2}P_{ph}P_{zh}-M^{2}P_{pz}(k^{2}-m^{2}_{h}+P_{hh}))+\]

\[
\xi k^{2}(P^{2}_{zh}(k^{2}-\xi P_{pp})+P^{2}_{ph}(k^{2}-\xi m^{2}_{z}-\xi P_{zz}))+\]

\[
M^{2}(k^{2}-\xi P_{pp})(k^{2}-m^{2}_{h}+P_{hh})(k^{2}-\xi m^{2}_{z}-\xi P_{zz})]/k^{2}d.\]

\[
G_{ph}=\xi [\xi P_{zg}(k^{2}(P_{ph}P_{zg}-P_{pg}P_{zh})-M^{2}P_{gh}P_{pz})+\xi M^{2}P_{pz}P_{zh}(k^{2}-\xi m_{z}^{2}+P_{gg})+\]

\[
M^{2}(k^{2}-\xi m^{2}_{z}-\xi P_{zz})(P_{ph}(k^{2}-\xi m^{2}_{z}+P_{gg})-P_{gh}P_{pg})]/k^{2}d.\]

\[
G_{zh}=\xi [\xi P_{pg}(k^{2}(P_{pg}P_{zh}-P_{ph}P_{zg})-M^{2}P_{gh}P_{pz})+\xi M^{2}P_{ph}P_{pz}(k^{2}-\xi m^{2}_{z}+P_{gg})-\]

\[
M^{2}(k^{2}-\xi P_{pp})(P_{gh}P_{zg}-P_{zh}(k^{2}-\xi m^{2}_{z}+P_{gg}))]/k^{2}d.\]

\[
G_{gh}=[\xi ^{2}(M^{2}P_{gh}P^{2}_{pz}-k^{2}P_{pz}(P_{ph}P_{zg}+P_{pg}P_{zh}))-\xi k^{2}(P_{zg}P_{zh}(k^{2}-\xi P_{pp})+\]

\[
P_{pg}P_{ph}(k^{2}-\xi m^{2}_{z}-\xi P_{zz}))-M^{2}P_{gh}(k^{2}-\xi P_{pp})(k^{2}-\xi m^{2}_{z}-\xi P_{zz})]/k^{2}d.\]

\[
G_{hh}=[\xi ^{2}P_{pz}(2k^{2}P_{pg}P_{zg}-M^{2}P_{pz}(k^{2}-\xi m^{2}_{z}+P_{gg}))+\xi k^{2}(P_{zg}^{2}(k^{2}-\xi P_{pp})+\]

\[
P_{pg}^{2}(k^{2}-\xi m^{2}_{z}-\xi P_{zz}))+M^{2}(k^{2}-\xi P_{pp})(k^{2}-\xi m^{2}_{z}+P_{gg})(k^{2}-\xi m^{2}_{z}-\xi P_{zz})]/k^{2}d.\]
The denominator is :
\[
d=\xi k^{2}[2P_{gh}P_{zg}P_{zh}-P_{zg}^{2}(k^{2}-m^{2}_{h}+P_{hh})-P^{2}_{zh}(k^{2}-\xi m^{2}_{z}+P_{gg})]+\]

\[
M^{2}(k^{2}-\xi m^{2}_{z}-\xi P_{zz})(P_{gh}^{2}-(k^{2}-m^{2}_{h}+P_{hh})(k^{2}-\xi m^{2}_{z}+P_{gg})).\]

The two other constraints due to the two last Ward identities give
two cubic polynomial equations. The first one ( 4.2 ) is quite simple
and reads: \( Mn_{pz}-i\xi m_{z}n_{pg}=0 \). In terms of the \( a_{ij} \) , we get :

\[
M(k^{2}(a_{46}(a_{59}a_{69}-a_{00}a_{56})+a_{49}(a_{56}a_{69}-a_{59}a_{66})+\]

\[
M^{2}a_{12}(a_{00}a_{66}-a_{69}^{2}))+i\xi m_{z}[k^{2}a_{59}(a_{46}a_{59}-a_{49}a_{56})+\]

\begin{equation}
\label{4.4}
M^{2}(a_{00}(a_{12}a_{56}-a_{33}a_{46})+a_{69}(a_{33}a_{49}-a_{12}a_{59}))]=0
\end{equation}

\noindent while in terms of self-energies we obtain :
\[
-M^{2}(k^{2}-m^{2}_{h})(k^{2}-\xi m^{2}_{z})(MP_{pz}+im_{z}P_{pg})-\]

\[
M[MP_{pz}((MP_{gg}+i\xi m_{z}P_{zg})(k^{2}-m^{2}_{h})+MP_{hh}(k^{2}-\xi m^{2}_{z}))-\]

\[
(k^{2}-\xi m^{2}_{z})P_{ph}(k^{2}P_{zh}+iMm_{z}P_{gh})-\]

\[
P_{pg}(iMm_{z}(P_{hh}(k^{2}-\xi m_{z}^{2})-\xi P_{zz}(k^{2}-m^{2}_{h}))+k^{2}P_{zg}(k^{2}-m^{2}_{h}))]+\]

\[
M^{2}P_{pz}(P_{gh}(MP_{gh}+i\xi m_{z}P_{zh})-P_{hh}(MP_{gg}+i\xi m_{z}P_{zg}))+\]

\[
P_{ph}(k^{2}P_{zh}(MP_{gg}+i\xi m_{z}P_{zg})-MP_{gh}(k^{2}P_{zg}-i\xi Mm_{z}P_{zz}))+\]

\begin{equation}
\label{4.5}
P_{pg}(MP_{hh}(k^{2}P_{zg}+i\xi Mm_{z}P_{zz})-k^{2}P_{zh}(MP_{gh}+i\xi m_{z}P_{zh}))=0.
\end{equation}

For the third Ward identity ( 4.3 ), the result is not so simple as
( 4.2 ); in terms of \( a_{ij} \) , we find

\[
Mk^{4}(a_{00}a_{46}^{2}+a_{66}a^{2}_{49}-2a_{46}a_{49}a_{69})+M^{3}k^{2}a_{11}(a^{2}_{69}-a_{00}a_{66})-\]

\[
Mk^{2}(k^{2}(a_{00}a^{2}_{56}+a_{66}a_{59}^{2}-2a_{56}a_{59}a_{69})+M^{2}a_{33}(a^{2}_{69}-a_{00}a_{66}))+\]

\[
2i\xi m_{z}k^{2}[k^{2}a_{49}(a_{49}a_{56}-a_{46}a_{59})+\]

\[
M^{2}(a_{00}(a_{12}a_{46}-a_{11}a_{56})+a_{69}(a_{11}a_{59}-a_{12}a_{49}))]+\]

\[
\xi ^{2}Mm_{z}^{2}k^{2}(a_{33}a_{49}^{2}-2a_{12}a_{49}a_{59}+a_{11}a_{59}^{2})+\]

\begin{equation}
\label{4.6}
\xi ^{2}M^{3}m^{2}_{z}a_{00}(a_{12}^{2}-a_{11}a_{33})=0.
\end{equation}

\noindent We don't write it in terms of the \( P_{ij} \) Green functions since
we can improve it thanks to the two other Ward identities (4.1 and
4.2). We briefly present the method. We first solve the equation (4.4),
i.e. we find \( a_{ij} \) as functions of the other \( a_{kl} \) ; \( a_{59} \) and \( a_{69} \) are not suitable
because they are irrationnal and complex functions. Secondly , we
substitute these \( a_{ij} \) in the equation ( 4.1) to get 7 equations , more
or less complicated , and not independent. After simplifications by
factors proportional to the determinant ( so different from zero ),
only the simplest equations are interesting; they are due to \( a_{00} \) , \( a_{33} \),
\( a_{56} \) and \( a_{66} \). It turns out that the first of them , written in functions
of the various \( P \) functions , gives only quadratic and cubic terms
in \( P \) , which is not really convenient for an eventual loop expansion.
Among the three other possibilities , the best choice we can do is
due to \( a_{56} \) for two reasons: it gives the most compact form and at one
loop order it is independent of the other results. Now we just have
to combine the result given by the equation (4.6) with the null identity
just derived to get in terms of the \( P \) functions:

\[
-M^{2}(k^{2}-m^{2}_{h})(Mm_{z}^{2}P_{gg}+2ik^{2}m_{z}P_{zg}+Mk^{2}P_{zz})\]

\[
M(M^{2}m_{z}^{2}P^{2}_{gh}+k^{2}(k^{2}-m^{2}_{h})P^{2}_{zg}-iMm_{z}P_{hh}(2k^{2}P_{zg}-iMm_{z}P_{gg})+\]

\[
2i\xi Mm_{z}(k^{2}-m^{2}_{h})(P_{pp}P_{zg}-P_{pg}P_{pz}+2ik^{2}Mm_{z}P_{gh}P_{zh}+k^{4}P_{zh}^{2}-\]

\[
M^{2}((k^{2}-m^{2}_{h})P_{gg}+k^{2}P_{hh})P_{zz})-2i\xi m_{z}(k^{2}P_{ph}(P_{ph}P_{zg}-P_{pg}P_{zh})+\]

\[
M^{2}(P_{hh}(P_{pg}P_{pz}-P_{pp}P_{zg})+P_{gh}(P_{pp}P_{zh}-P_{ph}P_{pz})))+\]

\[
M(k^{2}(P_{zh}(P_{gg}P_{zh}-P_{gh}P_{zg})+P_{zg}(P_{hh}P_{zg}-P_{gh}P_{zh}))+\]

\begin{equation}
\label{4.7}
M^{2}P_{zz}(P^{2}_{gh}-P_{gg}P_{hh}))=0.
\end{equation}

The last relations between the \( P \) Green functions come from the equation
( 2.7) ; from the relation concerning the photon , one easily gets:

\[
M^{2}(m^{2}_{h}-\xi m^{2}_{z})[(m_{h}^{2}-\xi m^{2}_{z})P_{ph}+P_{gg}P_{ph}-P_{gh}P_{pg}+\xi (P_{pz}P_{zh}-P_{ph}P_{zz})]\]

\[
+\xi [M^{2}(P_{gh}(P_{pg}P_{zz}-P_{pz}P_{zg})+P_{gg}(P_{pz}P_{zh}-P_{ph}P_{zz}))\]

\begin{equation}
\label{4.8}
+m^{2}_{h}P_{zg}(P_{ph}P_{zg}-P_{pg}P_{zh})]=0
\end{equation}
and from the relation with the \( Z \) boson, one obtains:
\[
M^{2}m^{2}_{h}(m^{2}_{h}-\xi m^{2}_{z})(m^{2}_{h}P_{zh}-iMm_{z}P_{gh})+\]

\[
M[\xi (m^{2}_{h}-\xi m^{2}_{z})(m_{h}^{2}P_{ph}(MP_{pz}-im_{z}P_{pg})-MP_{pp}(m^{2}_{h}P_{zh}-iMm_{z}P_{gh}))\]

\[
+m^{2}_{h}(m^{2}_{h}(MP_{gg}-i\xi m_{z}P_{zg})P_{zh}-MP_{gh}(m^{2}_{h}P_{zg}-i\xi Mm_{z}P_{zz}))]\]

\[
+\xi [m^{2}_{h}P_{zh}(P_{pg}(m^{2}_{h}P_{pg}-i\xi Mm_{z}P_{pz})-MP_{pp}(MP_{gg}-i\xi m_{z}P_{zg}))\]

\[
+M^{2}P_{gh}(P_{pz}(i\xi Mm_{z}P_{pz}-m^{2}_{h}P_{gg})+P_{pp}(m^{2}_{h}P_{zg}-i\xi Mm_{z}P_{zz}))\]

\begin{equation}
\label{4.9}
+m^{2}_{h}P_{ph}(P_{pz}(M^{2}P_{gg}-i\xi Mm_{z}P_{zg})-P_{pg}(m^{2}_{h}P_{zg}-i\xi Mm_{z}P_{zz}))]=0
\end{equation}
both equations at the energy of the \( A_{0} \) Higgs mass. In the Landau
gauge ( \( \xi =0 \) with our notations ), the two previous constraints turn
out to be very simple without cubic terms in \( P_{ij} \) and just reduce in
: 
\[
P_{ph}(m^{2}_{h}+P_{gg})=P_{gh}P_{pg},\]
and 
\[
P_{zh}(m^{2}_{h}+P_{gg})=P_{gh}(P_{zg}-iMm_{z}).\]

The first order expansion of all these results ( solutions and constraints
) are easy to derive. The solutions of the linear system become :
\[
G_{pp}=\xi /k^{2}\, \, ,\]

\[
G_{pz}=\xi ^{2}P_{pz}/(k^{2}(k^{2}-\xi m^{2}_{z}))\, \, ,\]

\[
G_{zz}=\xi (k^{2}-\xi m^{2}_{z}+\xi P_{zz})/(k^{2}-\xi m^{2}_{z})^{2}\, \, ,\]

\[
G_{pg}=-\xi P_{pg}/(k^{2}(k^{2}-\xi m^{2}_{z}))\, \, ,\]

\[
G_{zg}=-\xi P_{zg}/(k^{2}-\xi m^{2}_{z})^{2}\, \, ,\]

\[
G_{gg}=(-k^{2}+\xi m^{2}_{z}+P_{gg})/(k^{2}-m^{2}_{z})^{2}\, \, ,\]

\[
G_{ph}=-\xi P_{ph}/(k^{2}(k^{2}-m^{2}_{h}))\, \, ,\]

\[
G_{zh}=-\xi P_{zh}/((k^{2}-m^{2}_{h})(k^{2}-\xi m^{2}_{z}))\, \, ,\]

\[
G_{gh}=P_{gh}/((k^{2}-m^{2}_{h})(k^{2}-\xi m^{2}_{z}))\, \, ,\]

\[
G_{hh}=(-k^{2}+m^{2}_{h}+P_{hh})/(k^{2}-m^{2}_{h})^{2},\]
while the constraints derived from equations 4.1, 4.2 and 4.3 are
simplified as follows\( ^{2} \) : 
\[
P_{pp}=0\]

\[
MP_{pz}+im_{z}P_{pg}=0\]

\[
Mm_{z}^{2}P_{gg}+2ik^{2}m_{z}P_{zg}+k^{2}MP_{zz}=0.\]
At last, the constraints 2.8 and 2.9 read ( without any energy prescription
as we have proved in section 1-4 ):
\[
P_{ph}=0,\]
and

\[
k^{2}P_{zh}+im_{z}MP_{gh}=0.\]

The most striking difference between general and one loop results
is \( P_{pp}=0 \), which means the self -energy of the photon is only transverse
at one loop order. The second point to be stressed is a general feature:
everything is mixed beyond one loop, the CP odd Higgs included.

\section{Conclusion}

{In this study we have derived in a very general way the Schwinger
- Dyson equations for mixings between scalar and vector fields in
\( SU(2)*U(1) \) gauge models like the SM or the MSSM. We have found the solutions
of these equations and given some constraints induced by the BRS symmetry
they have to verify . Altough this study is rather technical, it can
be interesting for several reasons. For example, many authors use
the perturbation theory at more than one loop and can find useful
informations in our work. This is also true when one uses renormalization
group methods to improve first order results. An another interest
one can find in this paper is to check codes used to generate Feynman
diagrams and to calculate amplitudes. Indeed, since we only have considered
two point functions, all the formulae we have derived at one loop
order can be written in terms of B0 Veltman functions, favouring analytic
verifications. \\
\\
\bfseries \large Acknowledgments\par}

The author is grateful to members of the LPMT for very fruitful discussions.

\vspace{0.5cm}

{\bfseries \large \noindent References\par}

1 G. Altarelli, R. Barbieri and F. Caravaglios, CERN-TH 97-290, hep-ph
/ 9712368.

2 L. Baulieu and R. Coquereaux , Annals. Phys.140:163,1982 ; See also:
K. Aoki, Z. Hioki, R. Kawabe, M.Konuma and T.Muta., suppl. Prog. Theor.Phys.,
No 73,1982.

3 J.C. Collins, Renormalization, Cambridge University Press,1984 ;
R. Santos and A. Barroso, Phys. Rev. D 56, 5366 (1997).

4 J.F.Gunion, H.E.Haber, G.L.Kane and S.Dawson, The Higgs hunter's
guide, Addison - Wesley, Redwood City,1990.

5 M. Bohm, W.Hollik and H.Spiesberger, Fortschr.Phys.34 (1986) 687.

6 A. Dabelstein, Z.Phys.C67:495-512,1995.

7 J.A. Coarasa, D. Garcia, J. Guasch, R.A. Jimenez and J. Sola, UAB-FT
-397. hep-ph / 9607485.

\end{document}